\documentclass{ws-procs9x6}

\begin{document}

\title{
NEW FORMULATION OF STATISTICAL MECHANICS USING THERMAL PURE QUANTUM STATES
}

\author{Sho Sugiura$^*$ and Akira Shimizu}
\address{Department of Basic Science, 
University of Tokyo, 3-8-1 Komaba, Meguro, 
Tokyo 153-8902, Japan\\
$^*$E-mail: sugiura@asone.c.u-tokyo.ac.jp}

\begin{abstract}
We formulate statistical mechanics based on a pure quantum state, 
which we call a ``thermal pure quantum (TPQ) state''. 
A single TPQ state gives not only equilibrium values of mechanical variables, 
such as magnetization and correlation functions,  
but also those of genuine thermodynamic variables and thermodynamic functions, 
such as entropy and free energy. 
Among many possible TPQ states, we discuss 
the canonical TPQ state, 
the TPQ state whose temperature is specified. 
In the TPQ formulation of statistical mechanics, 
thermal fluctuations are completely included in 
quantum-mechanical fluctuations. 
As a consequence, TPQ states have much larger quantum entanglement 
than the equilibrium density operators of the ensemble formulation. 
We also show that the TPQ formulation is very useful in practical 
computations, by applying the formulation to a frustrated two-dimensional 
quantum spin system.
\end{abstract}


\bodymatter

\bigskip
\bigskip
\section{Introduction}\label{aba:sec1}
In quantum statistical mechanics, equilibrium states are conventionally described by mixed quantum states. 
By contrast, recent studies have shown the following fact \cite{SugitaJ, SugitaE, Popescu, Lebowitz, Reimann}.
Suppose that one prepares a pure quantum state as superposition of the energy eigenstates 
whose energies lie in the energy shell $[U-\Delta U,U+\Delta U]$ 
($U$: energy, $\Delta U$: energy width of $o(N)$). 
Then, almost every such pure state (measured by the Haar measure) gives the expectation values 
which are equal to those obtained from the microcanonical ensemble average 
with an exponentially small error, 
for any ``mechanical variables'' (See Sec.~\ref{Sec canTPQ}) such as magnetization and the correlation function. 
This result shows that a pure quantum state can represent a thermal equilibrium state. 
Motivated by this discovery, we generally call pure quantum states that give the correct equilibrium value for every mechanical variable thermal pure quantum (TPQ) states \cite{SS2012, SS2013}. 
 
However,``genuine thermodynamic variable'' such as temperature and the thermodynamic functions 
cannot be calculated as the expectation values of quantum-mechanical observables. 
In the ensemble formulation, they are related to the number of states. 
Therefore, one might think it impossible to obtain genuine thermodynamic variables from a {\em single} TPQ state. 
In this paper, however, we will show that genuine thermodynamic variables are related to 
the normalization constants of appropriate TPQ states. 
We present one example of such appropriate states, 
which we call the canonical TPQ state \cite{SS2013}. 

While the TPQ state of the previous works is specified by energy, 
the canonical TPQ state is specified not by energy but by temperature. 
We will show that the normalization constant of the canonical TPQ state gives the free energy. 
We also present another TPQ state specified by energy, 
whose normalization constant gives entropy. 
We call it the microcanonical TPQ state \cite{SS2012}. 
We show that the canonical TPQ state can be constructed efficiently from the microcanonical TPQ states.

These results establish a new formulation of statistical mechanics, 
which enables one to obtain 
all quantities of statistical-mechanical interest 
from a {\em single} realization of a TPQ state.
This formulation is 
not only interesting as fundamental physics 
but also advantageous 
in practical applications
because one needs only to construct a single pure state
by just multiplying the Hamiltonian matrix to a random vector. 

\section{Canonical TPQ State}\label{Sec canTPQ}
We consider a quantum system composed of $N$ sites (or particles). 
We assume that the dimension $D$ of its Hilbert subspace is finite.
[For particle systems, $D$ may be made finite by an appropriate truncation.]
We also assume that for this system the ensemble formulation gives correct results, which are consistent with thermodynamics in 
the thermodynamic limit, $N \rightarrow \infty$.
Here, we use the term ``thermodynamics'' 
in the sense of Refs.~\refcite{Callen,TD}.
[This means, for example, that the entropy function is concave.]
To exclude foolish operators such as $N^N \hat{H}$, we also assume that every mechanical variable is normalized as $\| \hat{A} \| \leq K N^m$ where $m$ is a constant of $o(N)$ and $K$ is a constant independent of $\hat{A}$ and $N$. 
We use quantities per site, 
e.g., $u \equiv E/N$ and $\hat{h} \equiv \hat{H}/N$.
The spectrum of $\hat{h}$ is assumed to be bounded, i.e., $e_{\rm min} \leq u \leq e_{\rm max}$.

The canonical TPQ state $|\beta, N \rangle$ 
is specified 
by the inverse temperature $\beta$ and $N$ 
(and possibly other variables such as magnetization, 
on which we do not explicitly write the dependence).
In order to generate it, take a random vector 
\begin{eqnarray}
| \psi_0 \rangle \equiv \sum_i c_i | i \rangle
\label{psi0}
\end{eqnarray}
from the whole Hilbert space.
Here, $\{ |i \rangle \}_i$ is an arbitrary orthonormal basis set of the whole Hilbert space 
and $\{c_i \}_i$ is a set of random complex numbers drawn uniformly from the $2D$ dimensional sphere, $\sum_i | c_i |^2=D$. 
Then, the canonical TPQ state is given by 
\begin{eqnarray}
|\beta, N \rangle 
\equiv
{\rm exp} \left[{-N \beta \hat{h} \over 2} \right]
| \psi_0 \rangle .
\end{eqnarray}
As we will see in the next two sections, 
it correctly gives both the thermodynamic functions and the equilibrium values of the mechanical variables. 

We notice that TPQ states are not the ``purification'' of mixed states 
(for details of purification, see Ref.~\refcite{NC}), 
because TPQ states are pure states in the $D$-dimensional Hilbert subspace,
i.e., they do not require an ancilla.

\section{Thermodynamic Functions and Genuine Thermodynamic Variables}
The free energy, which is one of thermodynamic functions, 
is obtained from the normalization constant of $|\beta, N \rangle $ as
\begin{equation}
f(\beta;N)
=
{1 \over \beta} \ln \langle \beta, N |\beta, N \rangle,
\end{equation}
where 
$f(\beta; N) \equiv -(1/\beta N) \ln Z(\beta, N)$ is the free energy density
[$Z(\beta, N)$ is the partition function].
Here, we write $(\beta; N)$ instead of $(\beta, N)$ in order to indicate that 
$f(\beta;N)$ converges to the $N$-independent one, $f(\beta)$. 

Using the random matrix theory and the generalized Markov inequality, 
the error probability is evaluated as 
\begin{eqnarray}
&&
{\rm P} \left( 
	\left|
	{\langle \beta,N | \beta,N \rangle
		\over	
	\exp[-N \beta f(1/\beta; N)]}
	-
	1
	\right|
	\geq \epsilon
	\right) 
\nonumber\\
&& \quad
\leq
{1 
\over 
\epsilon^2 \exp [2N\beta \{ f(1/2\beta ;N)-f(1/\beta ;N) \}]},
\label{Var<>}
\end{eqnarray}
where ${\rm P}(\cdot)$ is the probability that an event $\cdot$ happens.
%
Since $f(1/2\beta ;N)-f(1/\beta ;N)$ is positive 
and $\Theta(1)$ \cite{Theta} from thermodynamics\cite{TD}, 
the r.h.s. of inequality (\ref{Var<>}) is $\Theta({1 / \epsilon^2 \exp[N]})$. 
Therefore, a single realization of the canonical TPQ state almost always gives the correct thermodynamic function with an exponentially small error.
In another word,  
\begin{eqnarray}
{1 \over \beta} \ln \langle \beta, N |\beta, N \rangle 
\stackrel{P}{\to}
f(\beta)
\end{eqnarray}
where $\stackrel{P}{\to}$ denotes convergence in probability. 

All genuine thermodynamic variables
and any other thermodynamic functions
can be obtained from $f(\beta)$ by differentiation and 
the Legendre transformation. 

\section{Mechanical Variables}
In the previous section, we have shown that the canonical TPQ state correctly gives the free energy. 
The equilibrium values of all macroscopic quantities are derived from derivatives of the free energy. 
For mechanical variables, one can also obtain their equilibrium values
as the expectation values in the TPQ state.

The expectation value of a mechanical variable $\hat{A}$ in the canonical TPQ state 
\begin{eqnarray}
{\langle}\hat{A}{\rangle}^{\rm TPQ}_{\beta,N} 
\equiv 
{{\langle \beta,N |\hat{A}| \beta,N \rangle}
\over
{\langle \beta,N | \beta,N \rangle}}
\end{eqnarray} 
gives the equilibrium value with an exponentially small error. 
Like the ensemble average, the expectation value is useful in many practical applications. 

The squared average of the difference between this expectation value
and the canonical ensemble average 
\begin{eqnarray}
{\langle}\hat{A}{\rangle}^{\rm ens}_{\beta,N} 
\equiv
{
{\rm Tr} \, [e^{-N \beta \hat{h}} \hat{A}]
\over
Z(\beta,N) 
}
\end{eqnarray} 
is estimated as 
\begin{equation}
\overline{
( 
{\langle}\hat{A}{\rangle}^{\rm TPQ}_{\beta,N}
		-{\langle}\hat{A}{\rangle}^{\rm ens}_{\beta,N}
)^2 
}
\leq
{
\langle (\Delta \hat{A})^2 \rangle^{\rm ens}_{2\beta,N}
+
(\langle A \rangle^{\rm ens}_{2\beta,N} 
- \langle A \rangle^{\rm ens}_{\beta,N} )^2
\over \exp [2N\beta \{ f(1/2\beta ;N)-f(1/\beta ;N) \}]},
\label{Var<M>}
\end{equation}
where 
${\langle}(\Delta \hat{A})^2{\rangle}^{\rm ens}_{\beta,N}
{\equiv}{\langle}(\hat{A}-\langle{A}\rangle^{\rm ens}_{\beta,N})^2 
\rangle^{\rm ens}_{\beta,N}$. 
Using the generalized Markov inequality, we get an upper 
bound of the error probability as
\begin{eqnarray}
{\rm P} \left( 
	\left|
\langle \hat{A} \rangle^{\rm TPQ}_{\beta,N}
	-\langle \hat{A} \rangle^{\rm ens}_{\beta,N}
	\right|
	\geq \epsilon
	\right)
\leq
{1 \over \epsilon^2}
{
\langle (\Delta \hat{A})^2 \rangle^{\rm ens}_{2\beta,N}
+
(\langle A \rangle^{\rm ens}_{2\beta,N} 
- \langle A \rangle^{\rm ens}_{\beta,N} )^2
\over \exp [2N\beta \{ f(1/2\beta ;N)-f(1/\beta ;N) \}]}.
%
\label{P<}
\end{eqnarray}
Since $\| \hat{A} \| < K N^m$ (Sec.~\ref{Sec canTPQ}), 
the r.h.s. is $\Theta({N^m / \epsilon^2 \exp[N]})$. 
Therefore, 
a single realization of the canonical TPQ state almost always gives the correct equilibrium values of any mechanical variables with an exponentially small error.

We have shown that the equilibrium values of both mechanical and 
genuine thermodynamic variables 
are obtained from a single realization of the TPQ state. 
In this sense, we have established a new formulation of statistical mechanics based on a pure quantum state.

\section{A Numerical Application}\label{Sec Numerical}
Since our formulation requires only a single pure state 
for each equilibrium state, 
it is a powerful tool for practical applications. 
To illustrate this fact, 
we apply our formulation to a numerical computation in this section.

We present the result for spin-1/2 Kagome lattice Heisenberg antiferromagnet (KHA). 
This system is known to be hard to analyze because of frustration.
On the ground of the numerical diagonalization of small clusters up to N=18, it was suggested that the specific heat of KHA would have 
double peaks at low temperature \cite{Elser,N18,Sindzingre,Isoda}.

\begin{figure}[htbp]
\begin{center}
\includegraphics[width=0.9\linewidth]{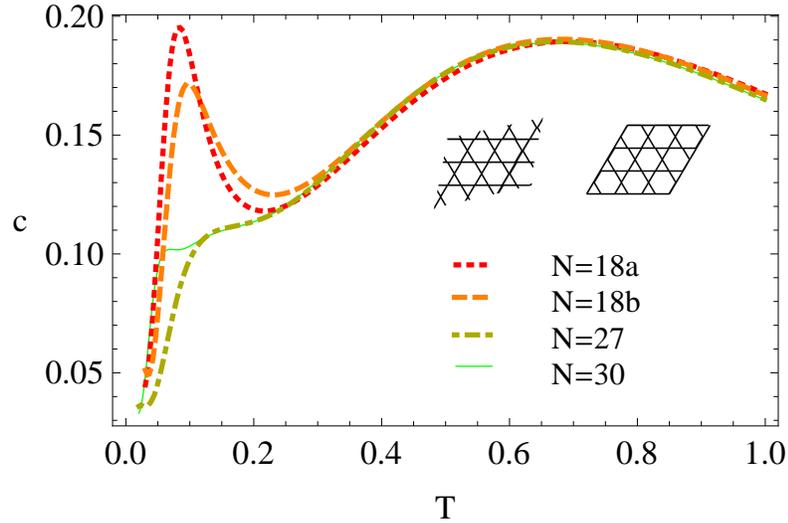}
\end{center}
\vspace{-6mm}
\caption{
$c$ vs. $T$ of the KHA. 
The shapes of clusters of $N=30$, $27$ and $18$a, $18$b
are shown in the right, left and in Ref.~\refcite{N18}, respectively.
}
\label{fig:c}
\end{figure}

In Fig~\ref{fig:c}, we show our results for the specific heat. 
[Some detail of the computation will be described in Sec.~\ref{Expansion}.]
The results for $N=18a$ and $b$ 
correspond to different shapes of the clusters. 
These results agree well with the previous results calculated by the numerical diagonalization, and show the double peaks.
However, 
the peak at lower temperature vanishes
for larger sizes, $N=27$ and $30$ [which cannot be treated by the numerical diagonalization].
We have obtained the results for these two clusters from a single realization of the canonical TPQ state. 
This suggests that the peak at lower temperature would be absent in the thermodynamic limit.

\begin{figure}[htbp]
\begin{center}
\includegraphics[width=0.9\linewidth]{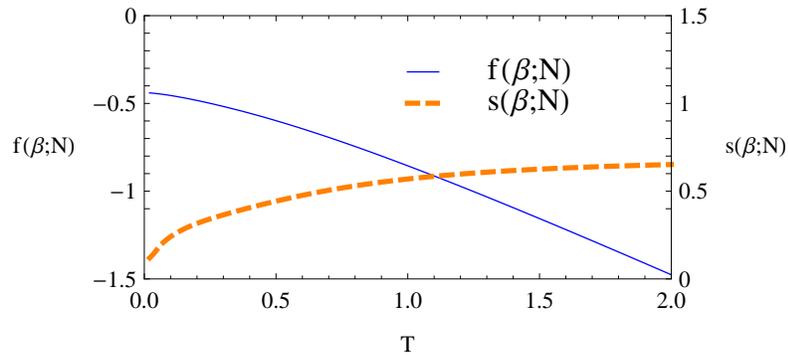}
\end{center}
\vspace{-6mm}
\caption{
$f$ and $s$ vs. $T$ for $N=30$. 
The shape of cluster is shown in Fig.~\ref{fig:c}.
}
\label{fig:fs}
\end{figure}

Another important result is the entropy density and the free energy density,
shown in Fig.~\ref{fig:fs}.
We observe that there remains $45$\% of the total entropy ($=N \ln 2$) 
at $T=0.2J$.
This is a consequence of strong frustration, which makes this system 
hard to analyze.

We emphasize again that these variables for $N=27$ and $30$ have been obtained from a single realization of the canonical TPQ state.
Moreover, recalling inequality~(\ref{P<}), we can estimate the probabilistic error of the result of the specific heat 
by using the result of the free energy density. 
The error is estimated to be less than 1\% down to $T=0.1J$. 
Thus, our new results for $N=27$ and $30$ are reliable enough 
for most purposes. 

\section{Microcanonical TPQ State} 
While we have generally defined TPQ states roughly in Sec.~\ref {aba:sec1},
we define it rigorously as follows.
%
%
When a state $|\psi{\rangle}$ is generated from some probability measure, 
it is called a TPQ state 
if 
\begin{eqnarray}
{\langle}\hat{A}{\rangle}^\psi_N
\stackrel{P}{\to}
{\langle}\hat{A}{\rangle}^{\rm ens}_N
\end{eqnarray}
uniformly for every mechanical variable $\hat{A}$ 
as $N \to \infty$.
Here, 
$
{\langle}\hat{A} \rangle^\psi_N
\equiv
{\langle}{\psi}|\hat{A}|{\psi}{\rangle}
/
\langle \psi | \psi \rangle
$,
$\langle{\cdot}\rangle^{\rm ens}_N$ is 
the ensemble average, and
`$\stackrel{P}{\to}$' denotes convergence in probability. 

This definition clearly shows that 
a single realization of the TPQ state for sufficiently large $N$ is enough to evaluate the equilibrium values of all {\em mechanical} variables.
Among such TPQ states are the random state 
in the energy shell\cite{SugitaJ, SugitaE, Popescu, Lebowitz, Reimann}
and the canonical TPQ state.
The latter has an additional special property that 
it also gives the equilibrium values of all 
{\em genuine thermodynamic} variables.
In this section, we present another TPQ state, 
called the microcanonical TPQ state, which also 
has this special property.

Starting from the random vector $|\psi_0 \rangle$ given by Eq.~(\ref{psi0}), 
the microcanonical TPQ state is defined by 
\begin{equation}
|k \rangle 
\equiv 
(l-\hat{h})^k |\psi_0 \rangle
\quad (k = 0, 1, 2, \cdots),
\label{mcTPQ}
\end{equation} 
where $l$ is an arbitrary constant s.t. $l \geq \{$maximum eigenvalue of $\hat{h} \}$.
The equilibrium value of the energy density is obtained by 
\begin{equation}
{\langle k|\hat{h} | k \rangle 
\over
\langle k | k \rangle }
\equiv 
u_k.
\end{equation}  
More generally, the equilibrium value of a mechanical variable $\hat{A}$ is obtained by 
\begin{equation}
{\langle k|\hat{A} | k \rangle 
\over
\langle k | k \rangle }.
\end{equation}  
We can show that this value, with increasing $N$, approaches the expectation value 
for the microcanonical ensemble of energy $N u_k$.
Thus, $|k \rangle$ satisfies the above condition for a TPQ state. 
Furthermore, 
it gives the entropy density $s(u)$ as
\begin{equation}
{1 \over N} \ln \langle k| k \rangle - {2k \over N} \ln (l-u_k)
\stackrel{P}{\to}
s(u_k).
\label{s=lnQ}
\end{equation}
%

Since the microcanonical TPQ state is generated by multiplying the polynomial of $\hat{h}$ to the random vector $|\psi_0 \rangle$, 
it can be generated easily, e.g., in a computer. 

\section{Expansion of the Canonical TPQ State} \label{Expansion}
The canonical TPQ state can be decomposed as the superposition of the microcanonical TPQ states. 
This decomposition 
enables one to perform numerical calculation efficiently.

We apply simple Taylor expansion to $\exp[N \beta(l-\hat{h})/2]$ as 
\begin{eqnarray}
|\beta,N \rangle 
&=&
e^{-N \beta l/ 2} 
\sum_{k=0}^{\infty} {(N \beta/2)^k \over k!} |k \rangle \\
&=&
e^{-N \beta l/ 2} 
\sum_{k=0}^{\infty} R_k |\psi_k \rangle.
\label{Phi=sumpsi_k}
\end{eqnarray}
where $|\psi_k \rangle \equiv |k \rangle /\sqrt{\langle k| k \rangle}$ 
is the normalized microcanonical TPQ state, 
and 
$R_k \equiv \sqrt{\langle k| k \rangle} (N \beta/2)^k / k!$.

Although this Taylor expansion is the sum of infinite terms, 
relevant $k$'s are not so many. 
To see the contribution of each $|k \rangle$ to $| \beta,N \rangle$, 
we focus on $R_k$ ($>0$). 
We can show that $R_k$ takes the maximum value for $k$ such that 
$u_k$ is closest to 
$\langle \beta,N |\hat{h}| \beta, N \rangle$. 
The values of $R_k$ for other $k$'s decay exponentially fast as the corresponding $u_k$ gets further from 
$\langle \beta,N |\hat{h}| \beta, N \rangle$. 
Thus, we can efficiently generate the canonical TPQ state 
from a small number of the microcanonical TPQ states, 
which can be numerically generated easily. 

The numerical results shown in Sec.~\ref{Sec Numerical} have been calculated using the above relation. 

\section{Quantum and Thermal Fluctuations}

To better understand the TPQ states, 
we now discuss the ``quantum fluctuation'' and ``thermal fluctuation''. 
For concreteness, we consider the canonical TPQ state $|\beta, N \rangle$ and 
the canonical density operator $\hat{\rho} = e^{- \beta N \hat{h}}/Z$.

In the ensemble formulation, it is often said that 
a fluctuation of a mechanical variable
$
\langle (\Delta \hat{A})^2 \rangle^{\rm ens} 
\equiv 
\langle (\hat{A}-\langle \hat{A} \rangle^{\rm ens})^2 \rangle^{\rm ens}
$
can be decomposed into 
the quantum fluctuation 
$\langle (\Delta \hat{A})^2 \rangle^{\rm ens}_{\rm q}$ 
and the thermal one
$\langle (\Delta \hat{A})^2 \rangle^{\rm ens}_{\rm t}$, i.e., 
\begin{eqnarray}
\langle (\Delta \hat{A})^2 \rangle^{\rm ens} 
=
\langle (\Delta \hat{A})^2 \rangle^{\rm ens}_{\rm q}+\langle (\Delta \hat{A})^2 \rangle^{\rm ens}_{\rm t}.
\label{F=Q+T}
\end{eqnarray}
The thermal fluctuation,
whose specific expression will be given below,
is conventionally interpreted as 
a result of mixing many quantum states to form $\hat{\rho}$,
\begin{equation}
\hat{\rho} = \sum_n (e^{-\beta N e_n}/Z) |n \rangle \langle n|,
\label{rho_in_n}\end{equation}
where $e_n$ and $|n \rangle$ are eigenvalue and eigenstate, respectively, of 
$\hat{h}$.
Consequently, it is conventionally concluded that the thermal fluctuation
of most mechanical variables 
does not vanish at any finite temperature.

In the TPQ formulation, by contrast, $| \beta,N \rangle$ is a pure quantum 
state and therefore does not have such ``thermal fluctuation'', i.e., 
$
\langle (\Delta \hat{A})^2 \rangle^{\rm TPQ}_{\rm t} 
=
0
$
at all temperature.
The TPQ state has only the quantum fluctuation, i.e., 
\begin{eqnarray}
\langle  (\Delta \hat{A})^2 \rangle^{\rm TPQ}
=
\langle  (\Delta \hat{A})^2 \rangle^{\rm TPQ}_{\rm q}
\equiv
\langle (\hat{A}-\langle \hat{A} \rangle^{\rm TPQ})^2 \rangle^{\rm TPQ}.
\end{eqnarray}
In other words, 
all fluctuations are included in the quantum fluctuation.

We have thus found that 
$\hat{\rho}$ and $| \beta,N \rangle$,
which represent the same equilibrium state,
give different values of the quantum and thermal fluctuations.
This does not lead to any contradiction in 
experimentally-observable quantities because
\begin{equation}
\langle (\Delta \hat{A})^2 \rangle^{\rm ens} 
=
\langle  (\Delta \hat{A})^2 \rangle^{\rm TPQ},
\end{equation}
which are the only observable quantities in the above discussion.
The quantum and thermal fluctuations,
$\langle (\Delta \hat{A})^2 \rangle^{\rm ens}_{\rm q}$ 
and
$\langle (\Delta \hat{A})^2 \rangle^{\rm ens}_{\rm t}$, 
are, separately, not observable quantities.
To see this, 
let us write them down explicitly.
We note that $\rho$ has the following form,
\begin{eqnarray}
\hat{\rho}
\equiv
\sum_{\lambda} w_{\lambda} | \lambda \rangle \langle \lambda |,
\label{DefRho}
\end{eqnarray}
where $\{w_{\lambda} \}_\lambda$ is a set of positive numbers such that 
$\sum_\lambda w_{\lambda} =1$,
and $\{ | \lambda \rangle \}_\lambda$ is some set of states
(which is $\{ | n \rangle \}_n$ in Eq.~(\ref{rho_in_n})).
In general, $\hat{A}$ fluctuates quantum-mechanically 
in each state $| \lambda \rangle$.
Hence, it may be reasonable to define
$\langle (\Delta \hat{A})^2 \rangle^{\rm ens}_{\rm q}$ 
as the average of the fluctuation 
$
\langle \lambda | 
( \hat{A} 
-
\langle \lambda | \hat{A} |\lambda \rangle
)^2
|\lambda \rangle
$
over $| \lambda \rangle$'s, i.e., 
\begin{equation}
\langle (\Delta \hat{A})^2 \rangle^{\rm ens}_{\rm q}
\equiv
\sum_\lambda w_\lambda
\langle \lambda | 
( \hat{A} 
-
\langle \lambda | \hat{A} |\lambda \rangle
)^2
|\lambda \rangle.
\label{QF}
\end{equation}
This and Eq.~(\ref{F=Q+T}) yield the thermal fluctuation as
\begin{equation}
\langle (\Delta \hat{A})^2 \rangle^{\rm ens}_{\rm t}
=
\sum_\lambda w_\lambda \langle \lambda | \hat{A} |\lambda \rangle^2
-
\left( 
\sum_\lambda w_\lambda \langle \lambda | \hat{A} |\lambda \rangle
\right)^2.
\label{TF}
\end{equation}
If we take 
$w_\lambda = e^{- \beta N e_n}/Z$ and $|\lambda \rangle = |n \rangle$,
we find that 
$\langle (\Delta \hat{A})^2 \rangle^{\rm ens}_{\rm t} > 0$ 
for most mechanical variables at finite temperature.

However, it is well-known that 
$| \lambda \rangle$'s in Eq.~(\ref{DefRho}) need not be 
orthogonal to each other \cite{NC}.
As a result, there are infinitely many possible choices of 
$\{ | \lambda \rangle \}_\lambda$ 
and $\{w_{\lambda} \}_\lambda$ 
for the same $\hat{\rho}$ \cite{NC}.
The experimentally-observable fluctuation
$\langle (\Delta \hat{A})^2 \rangle^{\rm ens}$ is invariant under the 
change of $\{w_{\lambda} \}_\lambda$ and $\{ | \lambda \rangle \}_\lambda$.
By contrast, 
both 
$\langle (\Delta \hat{A})^2 \rangle^{\rm ens}_{\rm q}$ 
and
$\langle (\Delta \hat{A})^2 \rangle^{\rm ens}_{\rm t}$
do alter under the 
change of $\{w_{\lambda} \}_\lambda$ and $\{ | \lambda \rangle \}_\lambda$.
This fact clearly shows that 
the quantum and thermal fluctuations 
are, separately, not experimentally-observable quantities.
In other words, they are, separately, metaphysical quantities.

It is instructive to consider
a classical mixture
\begin{eqnarray}
\hat{\rho}' 
\equiv
{1 \over R}
\sum_{r=1}^R
{
| \beta,N,r \rangle \langle \beta,N,r | 
\over
\langle \beta,N,r | \beta,N,r \rangle 
}
\end{eqnarray}
of many realizations 
$| \beta,N,1 \rangle, | \beta,N,2 \rangle, \cdots, | \beta,N,R \rangle$
of the canonical TPQ state.
Since each $| \beta,N,r \rangle$ represents the same equilibrium state, 
so does $\hat{\rho}'$.
If we define the quantum and thermal fluctuations in $\hat{\rho}'$ 
in the same way as Eqs.~(\ref{QF}) and (\ref{TF}), 
we find that the thermal fluctuation is exponentially small for all mechanical variables.
This shows that mixing many states does not necessarily give 
``thermal fluctuation''.
Since the thermal fluctuation in $\hat{\rho}'$ is negligible, 
we do not need to take an average over many relizations,
but only need to pick up a single realization.

\section{Entanglement} 
We have shown that the TPQ states and the density operators of the 
statistical ensembles give identical results for 
all quantities of statistical-mechanical interest.
That is, as far as one looks at macroscopic quantities, 
one cannot distinguish between these states. 
However, the TPQ states are pure quantum states while the 
density operators in the ensemble formulation (i.e., the Gibbs states)
are mixed states. 
Therefore, the situation changes when we look at entanglement. 
We discuss this point by studying entangelement of the 
microcanonical TPQ state.

To investigate entangelement of the TPQ state, 
we study its reduced density operator $\rho_q$ that is obtained 
by tracing out $N-q$ sites.
Its purity is defined by ${\rm Tr}(\rho_q^2)$. 
Since a TPQ state is a pure quantum state, 
this purity is a good measure of its entanglement.
The smaller the purity is, the  more entanglement the TPQ state has. 

\begin{figure}
\begin{center}
\includegraphics[width=0.9\linewidth]{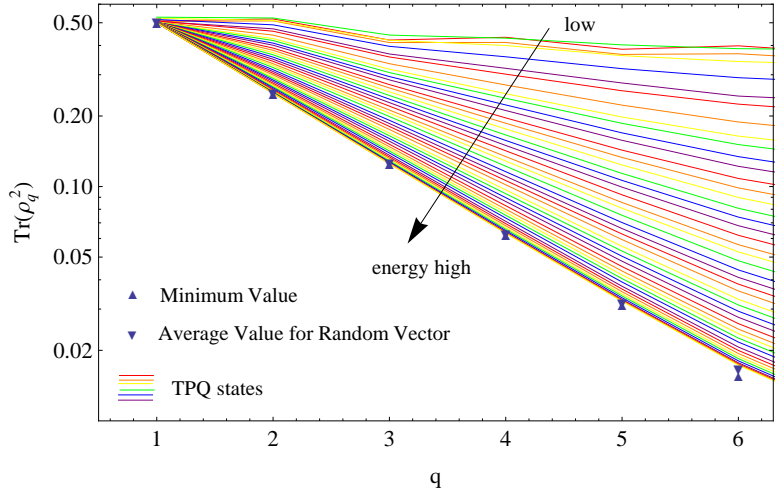}
\end{center}
\vspace{-6mm}
\caption{
Purity vs. $q$ of the 1D Heisenberg chain for $N=16$.
}
\label{Purity}
\end{figure}

In Fig.~\ref{Purity}, 
we plot the minimum value of the purity (triangles $\blacktriangle$) 
and the average value of the purity of the random vector $|\psi_0 \rangle$
(inverse triangles $\blacktriangledown$).
It is seen that $|\psi_0 \rangle$ has almost maximum 
(exponentially large) entanglement \cite{SS2005}. 
The lines are the purity of the microcanonical TPQ states 
with different values of the energy density.
It is seen that the TPQ states have exponentailly large entanglement,
and that the entanglement gets larger at higher energy, i.e., at higher temperature. 
This result is in marked contrast to entanglement of 
the density operator of the ensemble formulation,
because the latter has less entanglement at higher temperature.
[For example, with increasing temperature the canonical density operator approaches identity, which has no entanglement in any reasonable entanglement measure.]

However, this is not a contradiction but a natural consequence of 
the nature of entanglement.
The purity of $\rho_q$ is related to $N$-body correlation functions
of the TPQ state.
Such higher-order correlation functions represent microscopic details
of the TPQ state. 
Therefore, 
the great difference in entanglement between 
the TPQ states and the Gibbs states
indicates a great difference in microscopic details.
It is not surprising that 
such {\em microscopically} completely different states
give identical results for {\em macroscopic} quantities, 
and thus represent the same equilibrium state. 
 
\section{Conclusion}
In this paper, we have established a new formulation of statistical mechanics based on new TPQ states. 
A single realization of the TPQ state gives equilibrium values of all
mechanical and genuine thermodynamic variables and thermodynamic functions, with an exponentially small error. 
However, the TPQ states are completely different 
from the Gibbs states of the ensemble formulation. 
We have illustrated this fact by showing great difference of entanglement between them.
There are many possible TPQ states, such as the canonical TPQ state and the microcanonical one. 
The canonical TPQ state can be generated from the microcanonical ones, 
and the microcanonical ones can be obtained easily in a computer.
This fact makes the TPQ formulation advantageous in practical applications.


\begin{thebibliography}{9}
\bibitem{Theta}
For $f(N), g(N) > 0$, $f=\Theta(g)$ denotes that 
$^\exists a>0, ^\exists b >0, ^\exists N_0$ such that 
$a g(N) \leq f(N) \leq b g(N)$ for all $N >N_0$.
\end{thebibliography}
\end{document}